\documentclass[sigconf]{acmart}

\usepackage{amsmath}
\usepackage{soul}
\usepackage{color}
\usepackage{subfig}

\allowdisplaybreaks

\AtBeginDocument{%
  \providecommand\BibTeX{{%
    \normalfont B\kern-0.5em{\scshape i\kern-0.25em b}\kern-0.8em\TeX}}}

\copyrightyear{2021}
\acmYear{2021}
\setcopyright{acmcopyright}\acmConference[RecSys '21]{Fifteenth ACM Conference on Recommender Systems}{September 27-October 1, 2021}{Amsterdam, Netherlands}
\acmBooktitle{Fifteenth ACM Conference on Recommender Systems (RecSys '21), September 27-October 1, 2021, Amsterdam, Netherlands}
\acmPrice{15.00}
\acmDOI{10.1145/3460231.3474274}
\acmISBN{978-1-4503-8458-2/21/09}

\newtheorem{prop}{Proposition}
\newtheorem{definition}{Definition}

\begin{document}

\title{Debiased Explainable Pairwise Ranking from Implicit Feedback}


\author{Khalil Damak}
\email{khalil.damak@louisville.edu}
\affiliation{%
  \institution{Knowledge Discovery \& Web Mining Lab, Dept. of Computer Science \& Engineering, University of Louisville}
  \streetaddress{Louisville, KY 40292}
  \city{Louisville}
  \state{Kentucky}
  \country{USA}
  \postcode{40208}
}

\author{Sami Khenissi}
\email{sami.khenissi@louisville.edu}
\affiliation{%
  \institution{Knowledge Discovery \& Web Mining Lab, Dept. of Computer Science \& Engineering, University of Louisville}
  \streetaddress{Louisville, KY 40292}
  \city{Louisville}
  \state{Kentucky}
  \country{USA}
  \postcode{40208}
}

\author{Olfa Nasraoui}
\email{olfa.nasraoui@louisville.edu}
\affiliation{%
  \institution{Knowledge Discovery \& Web Mining Lab, Dept. of Computer Science \& Engineering, University of Louisville}
  \streetaddress{Louisville, KY 40292}
  \city{Louisville}
  \state{Kentucky}
  \country{USA}
  \postcode{40208}
}

\renewcommand{\shortauthors}{Damak et al.}

\begin{abstract}

Recent work in recommender systems has emphasized the importance of fairness, with a particular interest in bias and transparency, in addition to predictive accuracy. In this paper, we focus on the state of the art pairwise ranking model, Bayesian Personalized Ranking (BPR), which has previously been found to  outperform pointwise models in predictive accuracy, while also being able to handle implicit feedback. Specifically, we address two limitations of BPR: (1) BPR is a black box model that does not explain its outputs, thus limiting the user's trust in the recommendations, and the analyst's ability to scrutinize a model's outputs; and (2) BPR is vulnerable to exposure bias due to the data being Missing Not At Random (MNAR). This exposure bias usually translates into an unfairness against the least popular items because they risk being under-exposed by the recommender system.
In this work, we first propose a novel explainable loss function and a corresponding Matrix Factorization-based model called Explainable Bayesian Personalized Ranking (EBPR) that generates recommendations along with item-based explanations. Then, we theoretically quantify  additional exposure bias resulting from the explainability, and use it as a basis to propose an unbiased estimator for the ideal EBPR loss. The result is a ranking model that aptly captures both debiased and explainable user preferences. Finally, we perform an empirical study on three real-world datasets that demonstrate the advantages of our proposed models.

\end{abstract}



\maketitle

\section{Introduction}

Bayesian Personalized Ranking (BPR) is a state of the art pairwise ranking approach \cite{rendle2012bpr} that has recently received significant praise in the recommender systems community because of its capacity to rank implicit feedback data with high accuracy compared to pointwise models \cite{he2016vbpr}. Aiming to rank relevant items higher than irrelevant items, pairwise ranking recommender systems often assume that all non-interacted items as irrelevant. Hence, these systems rely on the assumption that implicit feedback data is Missing Completely At Random (MCAR), meaning that the items are equally likely to be observed by the users \cite{schnabel2016recommendations}, consequently any missing interaction is missing because the user chose not to interact with it. However, given the abundance of items on most e-commerce, entertainment, and other online platforms, it is safe to assume the impossibility of any user being exposed to \textit{all} the items. Thus, missing interactions should be considered Missing Not At Random (MNAR). This means that the user may have been exposed to part of the items, but chose not to interact with them, which can be a sign of irrelevance; and was not exposed to the rest of the items. This MNAR property is translated into an exposure bias. This type of bias is usually characterized by a bias against less popular items that have a lower propensity of being observed \cite{chen2020bias}.

Moreover, most accurate recommender systems tend to be black boxes that do not justify why or how an item was recommended to a user. This might engender unfairness issues if, for example, particularly inappropriate or offensive content gets recommended to a user. This kind of unfairness can be better diagnosed and mitigated with an explanation. In fact, it could be important for the user to know why or how the inappropriate item was recommended. For example, an Italian user might think that the movie recommendation ``The Godfather" is offensive because of the way it depicts, in an unfair  stereotypical way, a certain Italian community in the US. However, the explanation ``Because you liked the movie ``Scarface"" can be important in this case, because it clarifies that the movie recommendation was not tied to a community, but rather resulted from the user also liking another similar ``mafia" sub-genre movie. Furthermore, explanations have been shown to help users make more accurate decisions, which translates into an increased user satisfaction  \cite{bilgic2005explaining, abdollahi2017using}.
Bayesian Personalized Ranking \cite{rendle2012bpr} treats comparisons between any positive and negative items the same, regardless of which ones can or cannot be explained. 
Thus, while BPR aptly captures and models ranking based preference, it does not yet capture an \textit{explainable} preference. It is this \textit{explainable preference}, in addition to an unbiased preference ranking, that we seek to achieve in this work.
We thus propose models that address explainability \textit{and} exposure bias in pairwise ranking from implicit feedback and achieve the following contributions:

\begin{itemize}
    \item Proposing an explainable loss function based on the state of the art Bayesian Personalized Ranking (BPR) loss \cite{rendle2012bpr} along with a corresponding Matrix Factorization (MF)-based model called Explainable Bayesian Personalized Ranking (EBPR). To the extent of our knowledge, no work has introduced neighborhood-based explainability to pairwise ranking.
    \item Conducting a theoretical study of the additional exposure bias coming from the item-based explanations.
    \item Proposing an unbiased estimator for the ideal EBPR loss, called UEBPR, based on the Inverse Propensity Scoring (IPS) estimator \cite{saito2019unbiased}. To our knowledge, no prior work has tried to address the additional exposure bias that could result from neighborhood-based explainability.
    \item Performing an empirical study on three real-world datasets to compare the effectiveness of the proposed models, in terms of ranking, explainability, and both exposure and popularity debiasing.
    \item Investigating the properties of the proposed neighborhood based explainable models, revealing and explaining a desirable inherent popularity debiasing that is built into these models. This opens the path to a new family of future debiasing strategies, where the debiasing is rooted in an explainable neighborhood-based rationale.
\end{itemize}

In addition, we make our implementations of all the models presented in this paper available for reproducibility\footnote{\url{https://github.com/KhalilDMK/EBPR}}.

\section{Background}

In this section, we start by reviewing previous work on explainability and counteracting exposure bias in recommendation. While it is impossible to do justice to every past contribution with an exhaustive review, we try to focus on the most representative or related work. Then we review Bayesian Personalized Ranking (BPR).

\subsection{Explainability in Recommendation}

The types of explanations in recommendation have varied with the type of data used \cite{tintarev2007survey, bilgic2005explaining} 
%\hl{CITE (1) TINTAREV , (2) Bilgic and (3) herlocker}. 
Some explanations are content-based, meaning that they usually come from features. These were used in works that employed sentiment analysis on user reviews along with learned latent features to generate explanations in the form of user or item features \cite{zhang2014explicit}, textual sentences \cite{zhang2014explicit} or word clusters \cite{zhang2015incorporating}. Other research efforts used attention mechanisms to explain recommendations \cite{chen2017attentive, chen2018visually, seo2017interpretable, li2017neural}. The generated explanations are important regions in the textual \cite{seo2017interpretable} or image \cite{li2017neural, chen2017attentive, chen2018visually} inputs.
 %\cite{damak2019seer} proposed a hybrid deep learning architecture that generates a personalized segment as an explanation to a song recommendation by forward propagating segments of the song. 
Other methods relied on post-hoc approaches that try to extract explanations for the recommendations after they occur. For instance, \cite{rastegarpanah2017exploring} and \cite{cheng2019incorporating} use influence functions to determine the effect of each input interaction on the recommendation; while  \cite{damak2021sequence} proposed an approach that forward-propagates song segments through the trained recurrent neural network model to determine the most explanatory segment in a song recommendation.
In contrast to the above methods, some explainable recommender systems rely solely on feedback data such as ratings or interactions. Hence, they have the advantage of (1) accommodating collaborative filtering (CF) models and (2) not requiring any additional content or metadata to generate explanations for CF. These explanations tend to depend only on the rating data and they are mainly neighborhood-based, and can be either user-based or item-based \cite{herlocker2000explaining,abdollahi2017using}. Explanations can be obtained by using classical, inherently interpretable, user-based or item-based collaborative filtering techniques \cite{sarwar2001item, herlocker2000explaining} or by using model-based approaches. The latter are most related to our work. Among model-based approaches, Explainable Matrix Factorization (EMF) \cite{abdollahi2017using} pre-computes a user or item-based neighborhood style explainability matrix from the ratings, and then uses this matrix in a regularization term that is added to obtain an explainable recommendation reconstruction loss to guide the learning and yield explainable recommendations. This approach provides a simple and flexible way to add explainability to latent factor loss-based models to obtain a single integrated explainable model. It also has the advantage of not being a post-hoc approach, and hence not incurring the cost of learning two separate models, nor risking lack of fidelity from deviations between the explaining model and the predictive model. For all these reasons, EMF was later adopted in several works, such as \cite{coba2019personalised} which extended it and tried to improve the novelty of the recommendations; and in \cite{wang2018explainable} which modified the calculation of the explainability matrix by integrating the neighbors’ weights to improve performance. Other works used influence functions to generate neighborhood-based explanations. For instance, \cite{liu2019in2rec}  proposed a probabilistic factorization model, which employs an influence mechanism to evaluate the importance of the users’ historical data and present the most related users and items as explanations for the predicted rating.

\subsection{Exposure Bias in Recommendation}

 Bias in recommendation can be categorized into seven types \cite{chen2020bias} that occur within the various stages of the recommendation feedback loop \cite{khenissi2020theoretical, sun2019debiasing, jadidinejad2020using, nasraoui2016human} between the user, the data, and the model.
 %\hl{CITE some OLDER like Schnabel and also recent work from Himanipour, Burke, etc - see Sami's paper }. 
Among these categories, in the user-to-data phase, we find \textit{exposure bias}, which is the focus of our work in this paper. Exposure bias happens when users are only exposed to a portion of the items, and hence, unobserved interactions do not always represent negative preferences \cite{chen2020bias}.
The techniques that have been introduced to mitigate exposure bias, vary in whether they treat bias during the training or evaluation \cite{chen2020bias}. The common approach that is used in the evaluation phase incorporates an Inverse Propensity Scoring (IPS) modification of the ranking evaluation metrics, where more popular items are down-weighted and less popular items are up-weighted \cite{yang2018unbiased}. Exposure debiasing in training is usually achieved by considering the unobserved interactions as negatives with a certain confidence \cite{chen2020bias}. These methods differ in the way they define or approximate the confidence weight. One group of methods \cite{hu2008collaborative, devooght2015dynamic} considers a uniform weight for all negative items that is lower than one; while a second group \cite{pan2009mind, pan2008one} utilizes the user activity, for instance the number of interacted items, to weight the negative interactions; and a third group uses item popularity \cite{he2016fast, yu2017selection} and user-item similarity \cite{li2010improving} to achieve a similar goal. Recent work, \cite{saito2020unbiased} and \cite{saito2019unbiased}, proposed IPS-based unbiased estimators for the ideal pointwise and pairwise losses, respectively. In their experiments, they estimated the propensity of an interaction using the relative item popularity. On the other hand, \cite{khenissi2020modeling} proposed a regularization term that penalizes non-uniform exposure. Departing from the previously mentioned methods, other work proposed negative sampling processes in order to mitigate exposure bias. This negative sampling is usually done by exploiting side information such as social network information \cite{chen2019samwalker} or item-based knowledge graphs \cite{wang2020reinforced}. Another approach is to integrate the capacity to learn the exposure probability within the model \cite{liang2016modeling, chen2020fast, chen2019samwalker}, which in turn requires assumptions on the probability distribution of exposure. Finally, \cite{zhang2020large, ma2018entire, wen2020entire, bao2020gmcm} consider users’ sequential behavior to address exposure bias with multi-task learning.

\subsection{Bayesian Personalized Ranking for Pairwise Ranking}

The Bayesian Personalized Ranking (BPR) loss was introduced in \cite{rendle2012bpr} as the first loss that is ``optimized for ranking" in the implicit feedback pairwise ranking setting. In other words, it learns the users' preference of a positive item over a negative item. In this case, positive and negative items are those that the user has, respectively, interacted with and not interacted with. This is opposed to pointwise prediction, which can be seen as a predictive classification problem of the relevance of an item to a user. Pairwise ranking has received increasing attention and praise over the years from the recommender system community due to its high performance in top-N recommendation compared to pointwise ranking \cite{he2016vbpr}. The BPR objective function is defined as follows:

\begin{equation} \label{eq:bpr_loss}
    L_{BPR} = \frac{1}{|D|} \sum_{(u, i_{+}, i_{-}) \in D} - log \sigma (f_{\Omega}(u, i_{+}, i_{-})),
\end{equation}

where $D = \{(u, i_{+}, i_{-}) | u \epsilon U, i_{+} \epsilon I_{u}^{+}, i_{-} \epsilon I_{u}^{-}\}$ is the training data. $I_{u}^{+}$ is the set of positive (interacted) items by user $u$ and $I_{u}^{-}$ is the set of negative (non-interacted) items by user $u$ such that $I_{u}^{-} = I \setminus I_{u}^{+}$.
$f_{\Omega}$ is a hypothesis with parameters $\Omega$ that quantifies how much user $u$ prefers (following the order relation $>_{u}$ defined in \cite{rendle2012bpr}) item $i_{+}$ over item $i_{-}$, and $\sigma$ is the Sigmoid function.
%\begin{equation}
%    \sigma: x \mapsto \frac{1}{1 + e^{-x}}
%\end{equation}
When the BPR loss is applied to Matrix Factorization (MF) with the parameters $\Omega$ consisting of the respective user and item latent matrices $P \epsilon {\rm I\!R}^{n \times K}$ and $Q \epsilon {\rm I\!R}^{m \times K}$, the preference model is given by

\begin{equation}
    f_{\Omega}(u, i_{+}, i_{-}) = P_{u} \cdot Q_{i_{+}}^{T} - P_{u} \cdot Q_{i_{-}}^{T}. \label{eq:f_sigma}
\end{equation}

Applying the Sigmoid function to the output of the preference model yields the preference probability, which is the probability of user $u$ preferring item $i_{+}$ over item $i_{-}$: $P_{\Omega}(i_{+} >_{u} i_{-}) = P(i_{+} >_{u} i_{-} | \Omega) = \sigma (f_{\Omega}(u, i_{+}, i_{-}))$.
Note that in equation \ref{eq:bpr_loss}, as in the remainder of this paper, we omitted any regularization terms from the equations for simplicity, although we use L2 regularization in our implementation.

\section{Explainable Bayesian Personalized Ranking}

To the extent of our knowledge, no work has introduced neighborhood based explainability to pairwise ranking. More importantly, although neighborhood-based explainability can be expected to be vulnerable to exposure bias, there is a need to mitigate any additional exposure bias coming from the explainability. The BPR model learns to rank positive (interacted) items by a user higher than any negative (non-interacted) item. This objective treats comparisons between any positive and negative items the same, regardless of which ones can or cannot be explained based on any given style of explanation, for instance based on neighborhoods.
Thus, while BPR aptly captures and models a ranking based preference, it does not yet capture an \textit{explainable} preference.
In fact, as demonstrated in \cite{abdollahi2017using}, it is important to consider the interpretability of the items to the users, often referred to as \textit{explainability}, when learning a recommendation objective, and this can be computed based on readily available rating data, for instance from similar items.
Hence, given a definition for a measure of explainability  $E_{u i}$, of an item $i$ to a user $u$, our aim is to condition the BPR objective function to capture what we call \textit{explainable preference}. This means giving more importance to the explainable items that it is learning to rank higher, and less importance to the explainable items that it is learning to rank lower. In other words, if the objective function is learning to rank, for a user $u$, an item $i_{+}$ higher than an item $i_{-}$, then we would additionally want to give an even higher importance to this preference if it is also accompanied by a higher explainability $E_{u i_{+}}$ of item $i_{+}$ to user $u$ and a lower explainability $E_{u i_{-}}$ of item $i_{-}$ to user $u$.
We formulate this \textit{explainable preference} desiderata into a modified objective to obtain Explainable Bayesian Personalized Ranking (EBPR)  as follows:

\begin{definition}[Explainable Bayesian Personalized Ranking (EBPR) Objective Function]
Given an explainability matrix $E = (E_{u i})_{u=1..|U|, i=1..|I|} \in [0, 1]^{|U| \times |I|}$, where $E_{u i}$ is a measure of explainability of item $i$ to user $u$, the EBPR objective function is defined as

\begin{equation}
    L_{EBPR} = \frac{1}{|D|} \sum_{(u, i_{+}, i_{-}) \in D} - E_{u i_{+}} (1 - E_{u i_{-}}) log \sigma (f_{\Omega}(u, i_{+}, i_{-})). \label{eq:EBPR_Objective}
\end{equation}

\end{definition}

The intuition is to weight the contribution of an instance $(u, i_{+}, i_{-})$  into the loss by $E_{u i_{+}} (1 - E_{u i_{-}})$, in proportion to the degree that the positive item is considered to be more explainable and the negative item is considered less explainable. Hence, the higher the explainability $E_{u i_{+}}$ and the lower the explainability $E_{u i_{-}}$, the more the instance $(u, i_{+}, i_{-})$ will contribute to the learning. This also means that, when generating a recommendation list to a user $u$, the items ranked at the top of the list would be expected to have higher explainability than the items ranked lower in the list. Thus the multiplicative explainability term can be seen as one way to formulate an \textit{explainable preference} function, that is furthermore flexible, since any explainability score can be incorporated.

The latter objective function may seem counter-intuitive due to the fact that the loss increases when the explainability weighting term $E_{u i_{+}} (1 - E_{u i_{-}})$ increases. However, the model learns with the update equations regardless of the value of the loss. Hence, instead of trying to reduce the loss further when the explainability weighting term $E_{u i_{+}} (1 - E_{u i_{-}})$ increases, we aim to increase the \textit{contribution} of the instance $(u, i_{+}, i_{-})$ to the learning objective.
%, not sure if this is a good analogy plus it is actually a good idea so not giving away to reviewers :) \hlcyan{in much the same way in fact, as boosting in machine learning: the harder examples are given more weight in subsequent models - this actually works in the opposite away of above!}.
To gain a better insight, we derive the gradient used in the update equations of EBPR, with respect to the model parameters $\Omega$:

\begin{equation}
    \frac{\partial L_{EBPR}}{\partial \Omega} = \frac{-1}{|D|} \sum_{(u, i_{+}, i_{-}) \in D} E_{u i_{+}} (1 - E_{u i_{-}}) \frac{e^{-f_{\Omega}(u, i_{+}, i_{-})}}{1 + e^{-f_{\Omega}(u, i_{+}, i_{-})}} \frac{\partial f_{\Omega}(u, i_{+}, i_{-})}{\partial \Omega}.\label{eq:EBPR_update}
\end{equation}

From (\ref{eq:f_sigma}), we have

\begin{align*}
    \left.
  \begin{array}{r@{}l}
    \frac{\partial f_{\Omega}(u, i_{+}, i_{-})}{\partial \Omega}
  \end{array}
  \right. = \left\lbrace
  \begin{array}{l@{}l}
    Q_{i_{+} k} - Q_{i_{-} k} \;  \; \;if \; \Omega = P_{u k},\\
    P_{u k} \; \; \; \; \; \; \; \; \; \; \; \; \; \;\; if \; \Omega = Q_{i_{+} k},\\
    - P_{u k} \; \; \; \; \; \; \; \; \; \; \;\;\; if \; \Omega = Q_{i_{-} k},\\
    0 \; \; \; \; \; \; \; \; \; \; \; \; \; \; \; \; \; \;\; otherwise.
  \end{array}
  \right.
\end{align*}

The amplitude of the gradient with respect to parameter $\Omega$ is thus an increasing function of the explainability weighting factor $E_{u i_{+}} (1 - E_{u i_{-}})$ in a way that confirms the desired explainable preference aim. 
%\hlcyan{\textbf{Please comment this highlighted text but do not delete it - we will use in the future:} I think mathematically this measures the \textit{influence} of the data point on the prediction. This is actually cool. Influence methods are actually used for explanations and they have long history in robust statistical estimation, I believe they are obtained by second derivative of the loss function, so if you take the derivative of the gradient that gives an influence estimate. This is forcing the \textit{explainable} pairs to be more \textit{influential} on the model. I think we could probably reformulate it as a generic framework in the future. More on influence: https://christophm.github.io/interpretable-ml-book/influential.html and in fact may be I was right in my intuitive thinking about boosting, here Friedman uses the influence function to justify it: https://www.cc.gatech.edu/~zha/CS8803WST/boosting.pdf. This is like "both Explainability and Influence by Design"instead of using the instance's Influence for explainabilty as a product of the black box optimization (post-hoc without control on the model learning in the first place).
%This is really great if we can formulate it from the start as a new proposed framework and try to use similar theory as used by Friedman in justifying boosting, at some point he mentions influence functions}
For instance, in the extreme case where either the positive item is not explainable at all or the negative item is completely explainable, the update equation is zeroed out. Hence, no contribution will come from the corresponding instance to the learning. This is reasonable and desirable since the aforementioned case depicts a \textit{non} explainable preference, where either the positive item is not explainable or the negative item is explainable. Either case undermines the explainability of the preference.%, thus ranking of a positive item that has been observed over a negative item that has possibly not been observed.

\subsection{Explainability Matrix}

Various measures of explainability can be defined given the characterized order relation of an item $i$ being ``more explainable" than an item $j$ to a user $u$. The notion of explainability may depend on user or item metadata if using a content-based or hybrid approach. But in a purely collaborative filtering approach, such as in our case, it should be neighborhood-based as proposed in \cite{abdollahi2017using}, which further categorized the explanations as user-based or item-based. User-based explanations are based on user similarities and generate explanations in the form of ``this item was recommended because certain similar users liked it". Item-based explanations use item-similarities and generate explanations in the form ``the item was recommended because you liked similar items". We extend the idea of neighborhood-based explainability from \cite{abdollahi2017using} because it has shown success as an intuitive method for modifying loss-based recommendation models \cite{coba2019personalised, wang2018explainable}.
%\hlyellow{cite some papers that have cited Ref2 but not ours to maintain double blind, check google scholar and cite at least 4 other decent papers, including one by Symeonidis} \khalil{I was able to find two references}
Both item-based and user-based measures of explainability can be defined by relying solely on the interaction matrix (or rating matrix, depending on the type of feedback).
However, in this work, we focus only on item-based explanations which are expected to be more intuitive and informative to the user than user-based explanations. This is because the user knows the items that they interacted with but does not necessarily know their neighbors who have similar interactions with items. That said, a user-based explainability matrix can be similarly defined by applying the same strategy, described below, on the transpose of the interaction matrix.
We define the measure of \textit{explainability} $E_{u i}$ as the probability of user $u$ interacting with item $i$'s neighbors, as shown below.

\begin{definition}[Item-based explainability for Implicit Feedback]

\begin{equation}
    E_{u i} = P(Y_{u j} = 1 | j \in N_{i}^{\eta}), \label{eq:explainability}
\end{equation}

where $N_{i}^{\eta}$ is the neighborhood of item $i$ which is a set of item $i$'s $\eta$ most similar items given a similarity measure. $Y_{u i}$ is a Bernoulli random variable that takes value $1$ if user $u$ interacted with item $i$ and 0 otherwise:

\begin{align*}
    \left.
  \begin{array}{r@{}l}
    Y_{u i}
  \end{array}
  \right. = \left\lbrace
  \begin{array}{l@{}l}
    1 \; \; \;\;\;if \; i \in I_{u}^{+},\\
    0 \; \; \;\;\;otherwise.
  \end{array}
  \right.
\end{align*}

\end{definition}

The explainability $E_{u i}$ can also be reformulated as $E_{u i} = \frac{|N_{i}^{\eta} \cap I_{u}^{+}|}{\eta}$.
This means that for a specific item, the more neighboring items a given user has interacted with, the higher the explainability of that item will be to this user.
In our experiments, we use the Cosine similarity between items to generate the neighborhoods.

\subsubsection{Justifications for the Choice of Explainability}
In contrast to post-hoc explainability approaches, which generate explanations after the predictions have been made, our approach pre-computes explanation scores, then uses them to learn an explainable model. This leads to two advantages: (1) better transparency since there is no post-hoc model and (2) avoiding the heavy cost of post-hoc model training and explanation generation at prediction time.

Aiming toward \textit{transparency} is also why we chose to use \textit{neigh-borhood-based explainability}. More specifically, our aim is to explain recommendations using \textit{only} the input data used by the recommendation algorithm, and not any additional data that is not used to generate predictions. Consequently and because BPR uses no metadata, the explanations must be sourced from only the interaction data.

\subsection{Training Complexity of EBPR}

The complexity of learning the BPR model is $\mathcal{O}(|D|K)$, where $|D|$ is the size of the training data, and $K$ is the number of latent factors. 
This is because the complexity of forward and backward propagating an instance  stems from computing two dot products, which is  $\mathcal{O}(K)$. Considering that generating the explainability matrix can be done offline in the data pre-processing phase, no additional time complexity needs to be added to the training process of EBPR compared to BPR. 
That said, the  explainability matrix is computed only once, and the most significant part of the computation is computing the similarity values initially, which can be done very efficiently, owing to the sparsity of the interactions and the power law in the data distribution, allowing the use of sparse structures and locality sensitive hashing \cite{gionis1999similarity}.

\section{Exposure Bias in EBPR}

As proved in \cite{saito2019unbiased}, the estimator optimized in BPR is biased against the ideal pairwise loss. This is because the choice of the positive and negative items depends on the interaction random variable $Y_{u i}$ instead of the relevance. In fact, there is a discrepancy between interaction and relevance. Assuming that a relevant item is interacted implies that all non-interacted items are irrelevant, even if they were not exposed. This biases the BPR loss. Explainability too relies on the interaction random variable, hence amplifying this bias. 
To model exposure and relevance, we consider two Bernoulli random variables: $O_{u, i} \sim Ber(\theta_{u i})$, where $\theta_{u i} = P(O_{u i} = 1)$, represents the exposure propensity of item $i$ relative to user $u$; and $R_{u, i} \sim Ber(\gamma_{u i})$, where $\gamma_{u i} = P(R_{u i} = 1)$, represents the probability of item $i$ being relevant to user $u$. $O_{u, i}$ and $R_{u, i}$ represent, respectively, whether item $i$ is exposed or relevant to user $u$. We only know if user $u$ interacted with item $i$ when the item is both observed and relevant. In other words, $Y_{u i} = O_{u i} R_{u i}$ \cite{saito2019unbiased}. However, there could be relevant unobserved items that the user did not get a chance to observe in order to interact with.
To handle this issue, \cite{saito2019unbiased} proposed an Inverse Propensity Scoring (IPS) based estimator, as was done earlier for explicit feedback ratings in \cite{schnabel2016recommendations}, that is unbiased with respect to the ideal pairwise estimator. The latter is defined as follows.

\begin{definition}[Unbiased estimator for the ideal BPR loss]
\begin{equation}
    L_{BPR}^{unbiased} = \frac{1}{|U| |I|^{2}} \sum_{(u, i_{+}, i_{-}) \in U \times I \times I} - \frac{Y_{u i_{+}}}{\theta_{u i_{+}}} (1 - \frac{Y_{u i_{-}}}{\theta_{u i_{-}}}) log \sigma (f_{\Omega}(u, i_{+}, i_{-})).
\end{equation}
\end{definition}

Given that the explainability scores $E_{u i}$ also rely on the interaction random variable $Y_{u i}$, it is reasonable to suspect that the explainability weighting of the loss could introduce some additional exposure bias.
In fact, it would be ideal to use the relevance to define a more ideal explainability matrix as follows.

\begin{definition}[Ideal explainability matrix]

\begin{equation}
    E_{u i}^{ideal} = P(R_{u j} = 1 | j \in N_{i}^{\eta}).
\end{equation}

\end{definition}

This being done, we use the ideal explainability matrix to define the ideal EBPR loss as follows.

\begin{definition}[Ideal EBPR loss]

\begin{equation}
\begin{aligned}
    L_{EBPR}^{ideal} = & \frac{1}{|U| |I|^{2}} \sum_{(u, i_{+}, i_{-}) \in U \times I \times I} - \gamma_{u i_{+}} (1 - \gamma_{u i_{-}}) E_{u i_{+}}^{ideal} (1 - E_{u i_{-}}^{ideal})\\
    &\times log \sigma (f_{\Omega}(u, i_{+}, i_{-})).
    \end{aligned}
\end{equation}

\end{definition}

To quantify the additional bias, we compare the ideal EBPR loss to an IPS-based estimator similar to the one defined in Definition 3, but with explainability weighting. We call the latter estimator pUEBPR loss, where the ``pU" stands for \textit{partially unbiased}, and formulate it as follows.

\begin{definition}[Partially Unbiased Explainable BPR (pUEBPR) loss]

\begin{equation}
\begin{aligned}
    L_{pUEBPR} = & \frac{1}{|U| |I|^{2}} \sum_{(u, i_{+}, i_{-}) \in U \times I \times I} - \frac{Y_{u i_{+}}}{\theta_{u i_{+}}} (1 - \frac{Y_{u i_{-}}}{\theta_{u i_{-}}}) E_{u i_{+}} (1 - E_{u i_{-}})\\
    &\times log \sigma (f_{\Omega}(u, i_{+}, i_{-})).
\end{aligned}
\end{equation}

\end{definition}

The pUEBPR loss eliminates the initial exposure bias of BPR without taking into account the impact of adding explainability. Thus it is not a complete debiasing. However, as we will show below, this \textit{partial} debiasing loss will allow us to quantify the additional bias coming from adding the explainability weighting to BPR.
Next, we prove that the \textit{explainability} weighting in the EBPR loss introduces \textit{additional} exposure bias. Then we proceed to eliminate this additional bias in the next section.

\begin{prop}[Additional exposure bias from explainability weighting in EBPR] (proof is omitted)
The explainability weighting in the EBPR loss introduces additional non-zero exposure bias, given by

\begin{equation}
    Add\_Bias\_EBPR = \mathbb{E}[L_{pUEBPR}] - L_{EBPR}^{ideal}  \neq 0.
\end{equation}

\end{prop}

\section{Unbiased EBPR estimator}

We follow the same IPS-based methodology on the explainability weighting to propose an unbiased estimator for the ideal EBPR loss:

\begin{definition}[Unbiased EBPR (UEBPR) estimator]
\begin{equation}
\begin{aligned}
    L_{UEBPR} = &\frac{1}{|U| |I|^{2}} \sum_{(u, i_{+}, i_{-}) \in U \times I \times I} - \frac{Y_{u i_{+}}}{\theta_{u i_{+}}} (1 - \frac{Y_{u i_{-}}}{\theta_{u i_{-}}}) \frac{E_{u i_{+}}}{\theta_{u N_{i_{+}}^{\eta}}} (1 - \frac{E_{u i_{-}}}{\theta_{u N_{i_{-}}^{\eta}}})\\
    &\times log \sigma (f_{\Omega}(u, i_{+}, i_{-})),
\end{aligned}
\end{equation}

where $\theta_{u N_{i}^{\eta}} = P(O_{u j} = 1 | j \in N_{i}^{\eta})$ is the probability of user u being exposed to the neighbors of item $i$. $\theta_{u N_{i}^{\eta}}$ can also be considered as the item's \textit{neighborhood propensity} relative to user $u$.

\end{definition}

Now, we prove that this new UEPBR estimator is unbiased for the ideal EBPR loss in the following proposition.

\begin{prop}
The UEBPR estimator is unbiased for the ideal EBPR loss, meaning that

\begin{equation}
    \mathbb{E}[L_{UEBPR}] = L_{EBPR}^{ideal}.
\end{equation}

\end{prop}

\begin{proof}

\begin{align*}
    &\mathbb{E}[L_{UEBPR}] = %\mathbb{E}[\frac{1}{|U| \times |I|} \sum_{(u, i_{+}, i_{-}) \in U \times I \times I} - \frac{Y_{u i_{+}}}{\theta_{u i_{+}}} (1 - \frac{Y_{u i_{-}}}{\theta_{u i_{-}}}) \frac{E_{u i_{+}}}{\theta_{u N_{i_{+}}^{\eta}}} (1 - \frac{E_{u i_{-}}}{\theta_{u N_{i_{-}}^{\eta}}}) log \sigma (f_{\Omega}(u, i_{+}, i_{-}))]\\ 
    \frac{1}{|U| |I|^{2}} \sum_{(u, i_{+}, i_{-}) \in U \times I \times I} - \frac{\mathbb{E}[Y_{u i_{+}}]}{\theta_{u i_{+}}} (1 - \frac{\mathbb{E}[Y_{u i_{-}}]}{\theta_{u i_{-}}})\\
    &\times \frac{E_{u i_{+}}}{\theta_{u N_{i_{+}}^{\eta}}} (1 - \frac{E_{u i_{-}}}{\theta_{u N_{i_{-}}^{\eta}}}) log \sigma (f_{\Omega}(u, i_{+}, i_{-}))\\
    &=\frac{1}{|U| |I|^{2}} \sum_{(u, i_{+}, i_{-}) \in U \times I \times I} - \gamma_{u i_{+}} (1 - \gamma_{u i_{-}}) \frac{E_{u i_{+}}}{\theta_{u N_{i_{+}}^{\eta}}} (1 - \frac{E_{u i_{-}}}{\theta_{u N_{i_{-}}^{\eta}}})\\
    &\times log \sigma (f_{\Omega}(u, i_{+}, i_{-}))\\
    &=\frac{1}{|U| |I|^{2}} \sum_{(u, i_{+}, i_{-}) \in U \times I \times I} - \gamma_{u i_{+}} (1 - \gamma_{u i_{-}}) \frac{P(O_{u j} = 1, R_{u j} = 1 | j \in N_{i_{+}}^{\eta})}{\theta_{u N_{i_{+}}^{\eta}}}\\
    &\times (1 - \frac{P(O_{u j} = 1, R_{u j} = 1 | j \in N_{i_{-}}^{\eta})}{\theta_{u N_{i_{-}}^{\eta}}}) log \sigma (f_{\Omega}(u, i_{+}, i_{-}))\\
    %&=\frac{1}{|U| |I|^{2}} \sum_{(u, i_{+}, i_{-}) \in U \times I \times I} - \gamma_{u i_{+}} (1 - \gamma_{u i_{-}}) \frac{P(O_{u j} = 1 | j \in N_{i_{+}}^{\eta}) P(R_{u j} = 1 | j \in N_{i_{+}}^{\eta})}{\theta_{u N_{i_{+}}^{\eta}}} (1 - \frac{P(O_{u j} = 1 | j \in N_{i_{-}}^{\eta}) P(R_{u j} = 1 | j \in N_{i_{-}}^{\eta})}{\theta_{u N_{i_{-}}^{\eta}}})\\
    %&\times log \sigma (f_{\Omega}(u, i_{+}, i_{-}))\\
    &=\frac{1}{|U| |I|^{2}} \sum_{(u, i_{+}, i_{-}) \in U \times I \times I} - \gamma_{u i_{+}} (1 - \gamma_{u i_{-}}) \frac{\theta_{u N_{i_{+}}^{\eta}} E_{u i_{+}}^{ideal}}{\theta_{u N_{i_{+}}^{\eta}}} (1 - \frac{\theta_{u N_{i_{-}}^{\eta}} E_{u i_{-}}^{ideal}}{\theta_{u N_{i_{-}}^{\eta}}})\\
    &\times log \sigma (f_{\Omega}(u, i_{+}, i_{-}))
    %=\frac{1}{|U| \times |I|} \sum_{(u, i_{+}, i_{-}) \in U \times I \times I} - \gamma_{u i_{+}} (1 - \gamma_{u i_{-}}) E_{u i_{+}}^{ideal} (1 - E_{u i_{-}}^{ideal}) log \sigma (f_{\Omega}(u, i_{+}, i_{-}))
    = L_{EBPR}^{ideal}\qedhere
\end{align*}

\end{proof}

To get the last line, we assume conditional independence between exposure and relevance given the neighborhood, a much less restrictive (and thus more realistic) assumption than global independence.

\section{Experimental Evaluation}

We evaluate the impact of introducing explainability and counteracting exposure bias by tuning and then comparing the models described in Sections 2.3 -  5 in terms of ranking performance, explainability, and debiasing capabilities.
% that last comma often known as Oxford comma, it is recommended when listing a list of > 2 items with either and/or

\subsection{Data Used}

We use three datasets: The Movielens 100K \cite{harper2015movielens} (ml-100k), The Yahoo! R3 \cite{yahoo:r3dataset} (yahoo-r3) and the Last.FM 2K \cite{lastfm2kdataset, Cantador:RecSys2011} (lastfm-2k) datasets. These datasets consist of, respectively, 100K movie interactions, over 311K song interactions, and over 92K artist interactions. The interactions consist of either ratings or play counts, which were converted into binary interactions, regardless of their values. In fact, any rating or play count over the threshold of zero is considered a positive interaction. Then we filtered out users with less than 10 interactions in the lastfm-2k dataset to ensure enough training and evaluation samples for every user and reduce the data sparsity. The other two datasets similarly have at least 10 interactions per user. The datasets' properties are summarized in Table~\ref{tab:datasets}.

\begin{table}
\caption{Datasets used for evaluation. \label{tab:datasets} }
\begin{center}
{
\begin{tabular}{ l c c c c c}
\hline
Dataset & Task & Users & Items & Interactions & Sparsity\\ \hline \hline
ml-100k & Movie rec. & 943 & 1,682 & 100,000 & 93.6\%\\
yahoo-r3 & Song rec. & 15,400 & 1,000 & 311,704 & 97.9\%\\
lastfm-2k & Artist rec. & 1,874 & 17,612 & 92,780 & 99.7\%\\\hline
\end{tabular}
}
\end{center}
\end{table}

\subsection{Experimental Setting}

We follow the standard Leave-One-Out (LOO) procedure \cite{he2017neural, rendle2012bpr} that consists of considering the latest interaction of each user as a test item and comparing it to 100 randomly sampled negative items. In the training, we sample, at every epoch, one negative item for every positive user-item interaction. We implement ``BPR", ``UBPR", ``EBPR", ``pUEBPR" and ``UEBPR" and tune their hyperparameters on every dataset by comparing the averages over two replicates of 15 random hyperparameter configurations. We further split the training data into training and validation sets for the hyperparameter tuning. We consider the last interaction of every user from the training data along with 100 sampled negatives (disjoint from those in the test set) as a validation set. For each random hyperparameter configuration, we choose a value for the number of latent features, batch size and L2 regularization within the respective sets \{5, 10, 20, 50, 100\}, \{50, 100, 500\} and \{0, 0.00001, 0.001\}. We initially fixed the neighborhood size to 20 to ensure a fair comparison in terms of explainability metrics. However we will investigate the impact of neighborhood size later in Section \ref{Neighborhood_Section}. This being done, we then re-train every model on the merged train and validation sets with its best hyperparameter configuration for five replicates and report the average results on the test set. We also perform Tukey tests for pairwise comparison \citep{haynes2013tukey} to check the significance of the results. Note that, in our implementation of the unbiased models, namely UBPR, pUEBPR, and UEBPR, we only use positive and negative interaction pairs in the training to ensure that all models are trained on the exact same datasets, and truly assess the impact of every component in the loss. Also note that the goal of the experiments is to assess the impact of the added explainability and debiasing components on BPR. For this reason, we leave for future work, the task of comparing our algorithms to additional baselines.

\subsection{Evaluation Metrics}

We use Normalized Discounted Cumulative Gain ($NDCG@\mathcal{K}$) and Hit Ratio ($HR@\mathcal{K}$) for the ranking evaluation, and Mean Explainability Precision ($MEP@\mathcal{K}$) \cite{abdollahi2016explainable} and Weighted MEP ($WMEP@\mathcal{K}$) for the explainability evaluation. $MEP@\mathcal{K}$ measures the proportion of explainable items within the list of Top K recommendations, as follows

\begin{equation}
    MEP@\mathcal{K}(TopK) = \frac{1}{|U|} \sum_{u=1}^{|U|} \frac{|\{i \in TopK(u)\} \cap \{E_{u i} > 0\}|}{\mathcal{K}},
\end{equation}

where $TopK$ is the top $\mathcal{K}$ recommendation matrix in which every row represents the Top $\mathcal{K}$ recommendations of a user.
We further extend $MEP@\mathcal{K}$ to be able to weight the items' contributions to the numerator by their explainability values, since $MEP@\mathcal{K}$ rewards items that are considered to be explainable (i.e., with explainability score above a given threshold) in the same way, regardless of how different their explainability values are. Hence, we propose a weighted version of MEP that weights items' contributions by their explainability values. The\textit{ Weighted MEP (WMEP)} is given by

\begin{equation}
    WMEP@\mathcal{K}(TopK) = \frac{1}{|U|} \sum_{u=1}^{|U|} E_{u i} \frac{|\{i \in TopK(u)\} \cap \{E_{u i} > 0\}|}{\mathcal{K}}.
\end{equation}

Note that when training a model, we hide all test interactions when generating the explainability matrix to avoid any data leakage from the test set. Then, when evaluating the model on the test set, we generate an explainability matrix that considers all interactions to ensure an evaluation of the actual explainability of the test items to users.
Furthermore, we evaluate the popularity debiasing of the models in three aspects, namely Novelty, Popularity and Diversity. To evaluate the novelty of a model, we use Expected Free Discovery (EFD) \cite{vargas2011rank}, which is a measure of the ability of a system to recommend relevant long-tail items \cite{vargas2011rank}. EFD is defined as

\begin{equation}
    EFD@\mathcal{K}(TopK) = - \frac{1}{|U|} \sum_{u=1}^{|U|} \frac{1}{\mathcal{K}} \sum_{i \in TopK(u)} log_{2} \hat{\theta}_{u i}.
\end{equation}

Note that we use an estimator of the propensity $\hat{\theta}_{u i}$ to compute the popularity as we will see later in Section \ref{sec:propensity_estimation}. 
Next, to evaluate the popularity of the recommendations, we compute the average popularity at $\mathcal{K}$, using

\begin{equation}
    Avg\_Pop@\mathcal{K}(TopK) = \frac{1}{|U|} \sum_{u=1}^{|U|} \frac{1}{\mathcal{K}} \sum_{i \in TopK(u)} \hat{\theta}_{u i}.
\end{equation}

Finally, to evaluate recommendation diversity, we compute the Average Pairwise Similarity between the items in a top $\mathcal{K}$ recommendation list, which is given by \cite{vargas2011rank}

\begin{equation}
    Div@\mathcal{K}(TopK) = \frac{1}{|U|} \sum_{u=1}^{|U|} \frac{1}{\mathcal{K} (\mathcal{K} - 1)} \sum_{\substack{i, j \in TopK(u) \\ i < j}} sim(i, j),
\end{equation}

where $sim(i, j)$ is a measure of similarity between item $i$ and item $j$'s interaction vectors. In our experiments, we use the Cosine similarity.
All ranking and explainability metrics are computed at a cutoff $\mathcal{K}$ = 10 for Top 10 recommendation.

\subsection{Propensity Estimation} \label{sec:propensity_estimation}

Following \cite{saito2019unbiased}, we estimate the propensity of an item to a user by the relative item popularity of the item such that:

\begin{equation}
    \hat{\theta}_{u i} = \sqrt{\frac{\sum_{j = 1}^{|U|} Y_{j i}}{\max_{l \in I}\sum_{j = 1}^{|U|} Y_{j l}}}.
\end{equation}

The total propensity of item $i$ within its neighborhood can be defined as the average propensity of the items in the neighborhood\footnote{In our implementation, we ended up omitting the constant denominator in the sum as this yielded better results.}; i.e.,  $\hat{\theta}_{u N_{i}^{\eta}} = \frac{1}{\eta} \sum_{l \in N_{i}^{\eta}} \hat{\theta}_{u l}$.

\section{Results and Discussion}

\subsection{Overall Ranking and Explainability Results}

Table~\ref{tab:experimental_results} lists the results of all the models in terms of ranking performance and explainability.
Overall, for both the ml-100k and yahoo-r3 datasets, the explainable models EBPR and pUEBPR outperformed all the other models in terms of ranking performance and explainability for almost all the metrics. Moreover, whenever EBPR was not the best performer, it was still second to best.
On the lastfm-2k dataset, the non-explainable models (BPR and UBPR) reached better ranking performance than the explainable models (EBPR, pUEBPR and UEBPR). However, the explainable models were still the winners in terms of explainability (MEP and WMEP). Our interpretation of the exception in the lastfm-2k dataset, is that it is likely due to the extremely high sparsity of this dataset (99.7\%), which in turn impacts the similarity based computations to determine the neighborhoods used in computing the explainability values. This in turn degrades the learning of the explainable models due to the vanishing gradient problem. We will investigate this issue further in Section 7.5, where we will investigate the effect of the data sparsity on the learning of the explainable models.

\begin{table*}
\caption{Model comparison in terms of ranking performance and explainability on the three real interaction datasets that were described in Table 1. All evaluation metrics are computed at a cutoff $\mathcal{K}$=10 (Top 10) and reported as the averages over 5 replicates. The best results are in \textbf{bold} and second to best results are \underline{underlined}. A value with * is significantly higher than the next best value (p-value < 0.05). \label{tab:experimental_results} }
\begin{center}
{
\resizebox{\textwidth}{!}{\begin{tabular}{ l | c c c c | c c c c | c c c c}
\hline
Dataset & \multicolumn{4}{c|}{ml-100k} & \multicolumn{4}{c|}{yahoo-r3} & \multicolumn{4}{c}{lastfm-2k}\\ \hline
Model & NDCG & HR & MEP & WMEP & NDCG & HR & MEP & WMEP & NDCG & HR & MEP & WMEP\\ \hline \hline
BPR   & \underline{0.3807}* & \textbf{0.6625} & 0.9182* & 0.3467* & 0.3315* & 0.5466 & 0.8910* & 0.1594* & \textbf{0.7260}* & \textbf{0.9086}* & 0.2142 & 0.0452\\
UBPR & 0.3676* & 0.6401 & 0.9063* & 0.3342 & 0.3203 & 0.5422 & 0.8815 & 0.1562 & \underline{0.6613}* & \underline{0.8340}* & 0.2338 & 0.0468*\\\hline
EBPR & \textbf{0.3821}* & \underline{0.6568}* & \textbf{0.9314} & \textbf{0.3645}* & \textbf{0.3521} & \textbf{0.5674} & \textbf{0.9461}* & \textbf{0.1808}* & 0.6309* & 0.7876* & \textbf{0.2629}* & \textbf{0.0485}*\\
pUEBPR & 0.3648* & 0.6356* & \underline{0.9282}* & \underline{0.3595}* & \underline{0.3494}* & \underline{0.5662}* & \underline{0.9394}* & \underline{0.1778}* & 0.5938* & 0.7556* & \underline{0.2456}* & \underline{0.0471}*\\
UEBPR & 0.3542 & 0.6204 & 0.8986 & 0.3332 & 0.3421* & 0.5565* & 0.9234* & 0.1710* & 0.5567 & 0.7284 & 0.2349* & 0.0461\\
\hline
\end{tabular}}
}
\end{center}
\end{table*}

%\hlyellow{Leaving this comment here in case we decide to further explore in the future: can you control threshold values on purpose to induce different bias towards observing + vs - items ?}

\subsection{Advantages of using Explainability Weighting in the Learning Objective}

In order to demonstrate the advantages of the proposed explainability weighting in (\ref{eq:EBPR_Objective}), we compare EBPR to BPR and pUEBPR to UBPR because these models only differ by the explainability weighting of the loss. In both the ml-100k and yahoo-r3 datasets, going from BPR to EBPR almost always improves both the ranking and explainability performances. However, going from UBPR to pUEBPR improves the explainability but does not always improve the ranking performance. In fact, the ranking performance improves on the yahoo-r3 dataset but not on the ml-100k dataset. Nevertheless, we will see later, in Section 7.6, that pUEBPR outperforms UBPR on the ml-100k dataset when further tuning the neighborhood size. These results are somewhat surprising since while our initial aim was to improve the explainability of the recommended list, we ended up also gaining in ranking accuracy. In other words, explainability does not necessarily require sacrificing accuracy.

\subsection{Impact of Debiasing on Performance}

Contrary to what we noticed from the overall improved performance when adding explainability to any of the models, we notice a different trend in the accuracy when debiasing both models. In fact, on all three datasets, all the evaluation metrics decreased overall every time that debiasing was added: from EBPR to pUEBPR to UEBPR, and from BPR to UBPR. Hence, although the explainable models still perform  better overall than the non-explainable models, debiasing explainable models seems to be degrading the ranking performance. However, as the IPS weighting aimed to mitigate the exposure bias in the training phase, the evaluation sets still suffer from exposure bias. And given that the ranking metrics are based on the interaction, rather than relevance, they cannot properly quantify the benefits of the debiasing. To truly evaluate the impact of the exposure debiasing, we evaluate the models in terms of their capacity to capture the \textit{true relevance} which is only available in the yahoo-r3 dataset as described in the following subsection.

\subsection{Impact of Debiasing on Relevance Modeling}

\begin{table}
\caption{Model comparison in terms of ranking performance on the unbiased yahoo-r3 test set: Average results over 5 replicates. The best results are in \textbf{bold} and second to best are \underline{underlined}. A value with * is significantly higher than the next best value (p-value <0.05). \label{tab:relevance_results} }
\begin{center}
{
\begin{tabular}{ l | c c | c c c}
\hline
 & BPR & UBPR & EBPR & pUEBPR & UEBPR\\\hline\hline
NDCG@5 & 0.6140 & 0.6152 & 0.6178* & \textbf{0.6187} & \underline{0.6180}\\
MAP@5 & 0.4710 & 0.4727 & 0.4752* & \textbf{0.4764} & \underline{0.4756}\\
\hline
\end{tabular}
}
\end{center}
\end{table}

The yahoo-r3 dataset provides an unbiased test set, in which a subset of 5,400 users were provided 10 random songs to rate. The fact that the songs were chosen at random ensures that the test set is free of exposure bias, because all the rated songs have the same propensity of exposure. Thus, the ratings in the unbiased test set represent the true relevance of the items to the users. Hence, evaluating a model in terms of ranking performance on this test set reflects its capacity to capture the true relevance.
We re-train all the tuned models on the yahoo-r3 dataset, and evaluate it on the test set in terms of Mean Average Precision at cutoff 5 ($MAP@5$), and $NDCG@5$, where for both metrics, we assess the relevance of the top $\mathcal{K}$ predicted items for each user, given by their true rating-based ranking. We chose a cutoff of  5 because there are 10 test items per user. We summarize the results in Table~\ref{tab:relevance_results}.
Almost all the unbiased models performed better than their biased versions, except for pUEBPR which performed slightly better than UEBPR. This is probably due to the nature of the neighborhood propensity estimation. However, overall, the explainable and unbiased models, pUEBPR and UEBPR, were the best performers in terms of ranking performance in an unbiased evaluation setting. This demonstrates the impact of the loss debiasing in better accounting for the true relevance.

\subsection{Impact of Data Sparsity on Learning}

\begin{figure}
\centering
\includegraphics[width=0.4\textwidth]{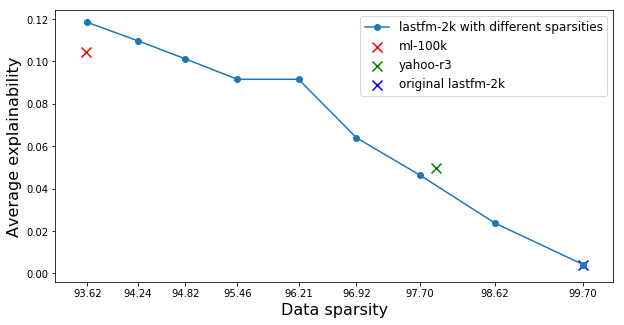}
\caption{Evolution of the average explainability with increasing sparsity of the lastfm-2k dataset. The average explainability values from the ml-100k and yahoo-r3 datasets are also shown for comparison. \label{fig:effect_of_sparsity}}
\end{figure}

In order to study the effect of the data sparsity on the performance of the explainable models, following our discussion in Section 7.1, we decided to explore the relationship between sparsity and explainability for the one data set (lastfm-2k) for which the performance trends differed. We do this by assessing the evolution of the explainability values from the explainability matrix, while gradually decreasing the sparsity of the dataset.
To reduce the data sparsity, we gradually, filtered out items with fewer than a certain threshold of interactions, namely 5, 10, 15, 20, 25, 30, 35 and 40 user interactions. %The resulting datasets, with their statistics, are summarized in Table~\ref{tab:datasets_reduced_sparsity}, along with the original dataset.
For each generated dataset, we compute the explainability matrix and calculate the average explainability value $E_{ui}$ in (\ref{eq:explainability}). We show the evolution of the average explainability with respect to the sparsity of the lastfm-2k dataset in  Fig.~\ref{fig:effect_of_sparsity}. We also show the average explainability values obtained from the ml-100k and yahoo-r3 datasets for comparison purposes.
The original lastfm-2k dataset has an average explainability of 0.0041 which is at least one order of magnitude lower than the average explainability values of 0.1043 and 0.0497 on the ml-100k and yahoo-r3 datasets, respectively. In the explainable models (EBPR, pUEBPR and UEBPR), the explainability values are multiplication factors in the update equations (\ref{eq:EBPR_update}). Hence, having explainability values that are close to 0 will cause the gradients to vanish and the learning to stall. Fig.~\ref{fig:effect_of_sparsity} shows a decreasing linear relationship between the explainability values and the data sparsity. Moreover, when reducing the lastfm-2k data sparsity to values near the respective sparsities of the ml-100k (93.6\%) and yahoo-r3 (97.9\%) datasets, we obtained average explainability values near those obtained from these two datasets. Thus, the data sparsity directly affects the scale of the explainability values. Higher data sparsity leads to lower explainability values and, consequently, a higher risk of vanishing gradients.
This confirms our suspicion, in Section 7.1, that the explainable models struggle with extremely sparse data due to the vanishing gradients problem.

%%%%%%%%%%%%%%%%%%%%%%%%%%%%%%%%%%%%%%%%%%%%  NOTE for FUTURE WORK: extremely important: add this as future goal, it ipens paths to explore explainability definitions in more detail and take into account sparsity challenges by counteracting vanishing gradients and other ways of defining/balancing the explainability etc in the objectove function like we did with oropensity debiasing, I suspect even propensity estimation itself may be affected by sparsity. %%%%%%%%%%%%%%%%%%%%%%%%%%%%%%%%%%%%%%%%%%%% 

%\hlyellow{New subsection from my email trying to justify why EBPR is superior, seems like we stumbled on a better debiasing that actually makes sense}

\subsection{Impact of Neighborhood Size on Performance} \label{Neighborhood_Section}

\begin{figure*}
\begin{minipage}{.23\linewidth}
\centering
\subfloat[]{\label{fig:effect_of_neighborhood_NDCG}\includegraphics[width=\textwidth]{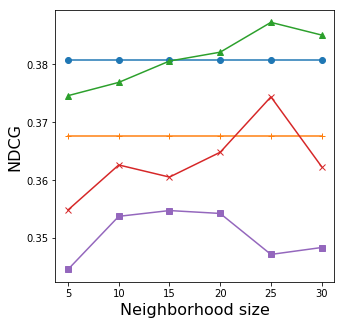}}
\end{minipage}
\begin{minipage}{.23\linewidth}
\centering
\subfloat[]{\label{fig:effect_of_neighborhood_HR}\includegraphics[width=\textwidth]{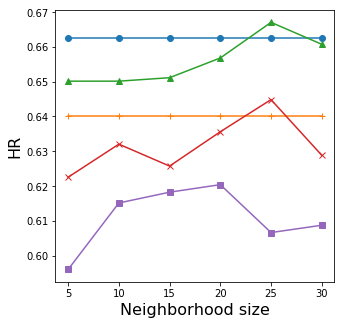}}
\end{minipage}
\begin{minipage}{.23\linewidth}
\centering
\subfloat[]{\label{fig:effect_of_neighborhood_MEP}\includegraphics[width=\textwidth]{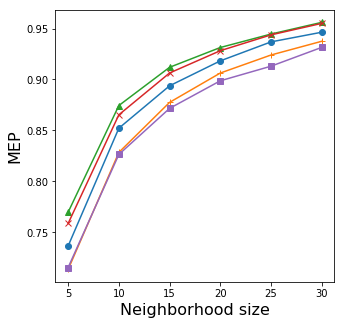}}
\end{minipage}
\begin{minipage}{.23\linewidth}
\centering
\subfloat[]{\label{fig:effect_of_neighborhood_WMEP}\includegraphics[width=\textwidth]{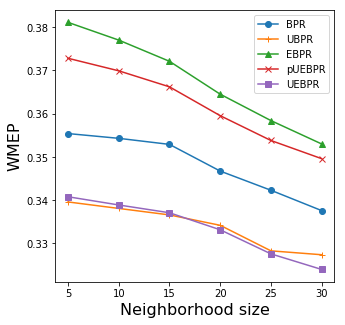}}
\end{minipage}\par\medskip
\caption{Evolution of NDCG@10, HR@10, MEP@10 and WMEP@10 with increasing neighborhood size on the ml-100k dataset. \label{fig:effect_of_neighborhood}}
\end{figure*}

The impact of the neighborhood size is two fold: First, the neighborhood size directly impacts the explainability values of items to users, which in turn impact the values of MEP and WMEP. For that reason, we used the same neighborhood size of 20 for all models in the hyperparameter tuning. Second, the explainability values, which depend on the neighbohood size, also impact the training of the explainable models EBPR, pUEBPR and UEBPR. Thus, to compare all models fairly in terms of ranking performance, the neighborhood size must be tuned for these explainable models.
In this section, we study the impact of the neighborhood size on the ranking accuracy and explainability. We vary the neighborhood size and re-train all the models in their optimal hyperparameter configurations. We show the results on the ml-100k dataset in Fig.~\ref{fig:effect_of_neighborhood}. We only show the results on the ml-100k dataset to avoid clutter and because we reached similar conclusions for the other two datasets.
As expected, the ranking accuracy (NDCG and HR) did not vary for the non-explainable models (BPR and UBPR) for the varying neighborhood sizes, contrarily to the explainable models (EBPR, pUEBPR and UEBPR), whose ranking prediction metrics showed different trends. EBPR and pUEBPR reached their highest ranking at a neighborhood size of 25, while UEBPR reached its maximum performance at 20. It is interesting to note that after tuning the neighborhood size, EBPR outperformed BPR and pUEBPR outperformed UBPR in both HR and NDCG which confirms our conclusions in Section 7.2, regarding the impact of the explainability weighting on the performance.
The explainability metrics show opposite trends with MEP increasing and WMEP decreasing when increasing the neighborhood size. This is due to the fact that larger neighborhood sizes lead to sparser neighborhoods and thus smaller explainability values, and the latter are used as a scale in the WMEP metric. Taking aside the trends, we see that the comparative performance of the models is somewhat consistent for different neighborhood sizes: Overall, EBPR yields the best explainability performance for all neighborhood sizes, followed by pUEBPR.

\begin{table*}
\caption{Model comparison in terms of Novelty (EFD), Popularity (Avg\_Pop) and Diversity (Div) on the three datasets. All evaluation metrics are computed at a cutoff $\mathcal{K}$=10 (Top 10) and reported as the averages over 5 replicates. The best results are in \textbf{bold} and second to best results are \underline{underlined}. HB means the higher the better and LB means the lower the better. Any value with * is significantly higher than the next best value (p-value < 0.05). \label{tab:popularity_bias_results} }
\begin{center}
{
\resizebox{\textwidth}{!}{\begin{tabular}{ l | c c c | c c c | c c c}
\hline
Dataset & \multicolumn{3}{c|}{ml-100k} & \multicolumn{3}{c|}{yahoo-r3} & \multicolumn{3}{c}{lastfm-2k}\\ \hline
Model & EFD (HB) & Avg\_Pop (LB) & Div (LB) & EFD (HB) & Avg\_Pop (LB) & Div (LB) & EFD (HB) & Avg\_Pop (LB) & Div (LB)\\\hline \hline
BPR & 1.2029 & 0.4739 & 0.2675 & 1.7681 & 0.3460 & 0.0811* & 2.7714 & 0.2000 & 0.0184\\
UBPR & \underline{1.3445}* & \underline{0.4397}* & \underline{0.2497}* & \underline{1.8157} & 0.3348* & \textbf{0.0789}* & 3.1049* & 0.1714* & 0.0163*\\\hline
EBPR & 1.2160 & 0.4677* & 0.2650* & 1.7682 & 0.3442 & 0.0844 & \textbf{3.4056}* & \underline{0.1521}* & 0.0146*\\
pUEBPR & 1.2939* & 0.4491* & 0.2587* & 1.8148* & \underline{0.3341} & 0.0822* & 3.3446 & 0.1531* & \underline{0.0137}*\\
UEBPR & \textbf{1.4699}* & \textbf{0.4127}* & \textbf{0.2414}* & \textbf{1.8716}* & \textbf{0.3222}* & \underline{0.0800}* & \underline{3.3843}* & \textbf{0.1478} & \textbf{0.0130}*\\
\hline
\end{tabular}}
}
\end{center}
\end{table*}

\subsection{Explainability as Debiasing or \textit{Explainable Debiasing}}

EBPR's superior accuracy with no apparent tradeoff with explainability suggests an inherent popularity debiasing mechanism that is a byproduct of adding explainability. This is certainly possible because the explainability term $E_{ui_{+}}(1 - E_{ui_{-}})$, when multiplied into the ranking accuracy loss, captures finer detail about an item's rating from the item's neighbors in addition to the item's own rating. This term has therefore ended up counteracting the bias of very popular items by relying on their neighborhoods.
In fact, the explainability weighting term is expected to pull very popular items down, similarly to propensity debiasing. However what the proposed explainability term, ends up doing, in contrast to propensity debiasing, is succeeding in the estimation of propensity, more accurately and in a local way, namely by using the neighborhood around each item, and not solely the item itself. The advantage of the explainability term is also that it takes into account the local neighborhood to provide a rationale for both positive and negative interactions. Indeed the explainability score is not only providing intuitive quantitative explanation scores for output predictions, but also providing a rationale for debiasing, effectively providing what can be considered an \textit{explainable local debiasing} strategy for each item. Next, we investigate this powerful idea for local explainable propensity estimation by evaluating and comparing the models in terms of Novelty (EFD), Popularity (Avg\_Pop) and Diversity (Div). We summarize our results in Table~\ref{tab:popularity_bias_results}.
For all datasets and for almost all evaluation metrics, the explainable model EBPR outperformed the vanilla BPR, thus supporting our aforementioned claims of popularity debiaing with explainability weighting.
Moreover, adding the exposure debiasing (moving from BPR to UBPR or moving from EBPR to pUEBPR then UEBPR) almost always improves the popularity bias metrics. This demonstrates a relationship between exposure bias and popularity bias where mitigating the former consequently mitigates the latter.
%%%%%%%%%%%%%% FUTURE: show this theoretically! also the word "causal" is too strong
Finally, UEBPR showed the best popularity debiasing overall on all the datasets. The considerably high debiasing performance of UEBPR is likely due to its down-weighting of the items with popular \textit{neighborhoods}, in addition to the popular items, hence allowing the less popular items to be discovered. We plan to investigate this further in future work.
%%%%%%%%%%%% FUTURE: show this theoretically (show the neighborhood impact)

\section{Conclusion}

We proposed a novel explainable pairwise ranking loss with a corresponding MF-based model called Explainable Bayesian Personalized Ranking. We theoretically quantified the additional exposure bias resulting from the explainability, and proposed an IPS-based unbiased estimator for the ideal loss. We tested our proposed approaches on three recommendation tasks and presented an extensive discussion about the advantages of the proposed explainability extension; as well as the impact of the debiasing, for varying data sparsities and varying neighborhood sizes. Finally, we studied the popularity-debiasing properties of the proposed methods in terms of Novelty, Popularity, and Diversity; and unveiled an inherent popularity debiasing stemming from the neighborhood interactions. Our findings are informative and motivate further research because our proposed EBPR model yielded the best performance overall with no significant trade-off between explainability and accuracy. Moreover, we showed how combining explainability and exposure debiaing yields powerful popularity debiasing through the proposed UEBPR loss. Finally, our results point towards EBPR and pUEBPR being the top performers that offer the best tradeoff between accuracy, explainability and debiasing capacity. However, despite their competitive performance, our proposed approaches may suffer from the vanishing gradient problem in extremely sparse settings.
%In the future, we plan to further investigate theoretically the inherent debiasing properties of our proposed explainability weighting and investigate how our unbiased estimators approximate the true losses in a semi-synthetic setting in which the relevance is estimated.

\begin{acks}
This work was supported in part by National Science Foundation grant IIS-1549981.
\end{acks}

\bibliographystyle{ACM-Reference-Format}
\bibliography{sample-base}

\end{document}